\documentclass[preprint,aps,floatfix]{revtex4-1}
\usepackage{hyperref}
\everymath={\displaystyle}
\usepackage{young}
\usepackage{rotating}
\usepackage{longtable}
\def\1{\mbox{l\hspace{-0.53em}1}}

\usepackage{multirow}
\newlength{\AccoHaut}

\begin{document}
\title{Hidden bottom pentaquark in the SU(5) 
version of the flavor-spin model}
\author{Fl. Stancu\thanks{e-mail : fstancu@ulg.ac.be}}
\address{Universit\'{e} de Li\`ege, Institut de Physique B.5, Sart Tilman,
B-4000 Li\`ege 1, Belgium}

\date{\today}
\everymath={\displaystyle}

\begin{abstract}

We generalize to five distinct flavors the flavor-spin hyperfine interaction introduced 
previously for four flavors and used in the study of  $uudc\overline{c}$ pentaquark. 
As a particular case here
we study the lowest states of the pentaquark $uudb\overline{b}$,
of either positive or  negative parity,  
in a constituent quark model with linear confinement 
and the presently extended hyperfine interaction.
The positive parity states have one unit of angular momentum located in the subsystem of four quarks
and are described by translationally invariant states of orbital permutation symmetry $[31]_O$
which requires the configuration $s^3 p$. The negative parity states are described by the 
configuration $s^4$ of permutation symmetry $[4]_O$.   
We show that the lowest state has the quantum numbers  $J^P$ = $1/2^+$ or $3/2^+$ and I = $1/2$ and is located  
below  the $\Sigma_b B$ threshold by  - 132 MeV.
We present a comparison between the spectra of  $uudc\overline{c}$ and $uudb\overline{b}$ pentaquarks. 
\end{abstract}

\maketitle

\vspace{1cm}

\section{Introduction}

In 2015 the LHCb Collaboration reported the existence of two pentaquark-like  
resonances  named $P^+_c!4380)$ and $P^+_c(4450)$ in the $\Lambda^0_b \rightarrow J/\psi p K^-$ decay \cite{Aaij:2015tga}.
Due to the $J/\psi p$ component 
these structures were interpreted as hidden charm pentaquarks of flavor  content  $uudc\overline{c}$.

In 2019 the LHCb collaboration updated the analysis of 
the $\Lambda^0_b \rightarrow J/\psi p K^-$ decay \cite{Aaij:2019vzc} and   
reported the existence of three narrow structures 
named  $P^+_c(4312)$, $P^+_c(4440)$ and $P^+_c(4457)$ where  $P^+_c(4312)$ was entirely new.
The other two resonances replaced the previous $P^+_c(4450)$.
The broad  $P^+_c(4380)$ resonance awaits confirmation.

In 2021 following the observation of hidden charm pentaquarks 
the LHCb collaboration reported a new hadronic exotic state named $P^0_{cs}(4459)$ in the invariant mass
distribution of the  $\Xi^{-}_b \rightarrow J/\psi \Lambda K^-$ decay \cite{LHCb:2020jpq}.
Because of the quark content of $J/\psi$ and  $\Lambda$
this structure is a candidate for a strange hidden charm pentaquark of flavor content $udsc\overline{c}$.
Very recently the LHCb collaboration detected another strange hidden charm pentaquark with 
preferred quantum numbers $J^P = 1/2^-$ named ${P_{cs}(4338)}$  \cite{LHCb:2022}.

The new name of $P^+_c(4312)$, recently proposed by the LHCb collaboration
 is ${P^N_{\psi}(4312)}^+$ and the new name of  ${P^0_{cs}(4459)}$ 
is ${P^{\Lambda}_{{\psi}s}(4459)}$  
\cite{Gershon:2022xnn}.

The 2019  LHCb observation has triggered many theoretical interpretations.
Although observed in the $J/\psi p$ channel,
the proximity of the mass of  
$P^+_c(4312)$ 
and of the masses of  $P^+_c(4440)$ and $P^+_c(4457)$ 
to the respective thresholds favored the molecular scenario
 \cite{Guo:2019kdc,Guo:2019fdo,Xiao:2019mst,Xiao:2019aya,Shimizu:2019ptd,Lin:2019qiv,
Liu:2019tjn,Meng:2019ilv,Wu:2019rog,Valderrama:2019chc,Du:2019pij,Wang:2019spc,Xu:2020gjl,Chen:2020pac,Burns:2019iih,
Fernandez-Ramirez:2019koa}.

The 2019 LHCb pentaquarks have also been analyzed in compact pentaquark models based on the chromomagnetic interaction 
of the one gluon exchange model,
with quark/antiquark correlations \cite{Ali:2019npk} or without correlations \cite{Weng:2019ynv,Cheng:2019obk}.
In both cases the lowest state has negative parity.

The spectrum of positive and negative parity states of the $uud c \bar c$ pentaquarks 
has also been studied in a constituent quark model with an SU(4)
flavor-spin hyperfine interaction  \cite {Stancu:2019qga} which is a generalization of
the  model of Ref. \cite{Glozman:1995fu} based on SU(3).  
The merit of the flavor-spin (FS) model   
is that it reproduces the correct ordering of positive and negative parity
states of both nonstrange and strange baryons \cite{Glozman:1995fu,Glozman:1996wq,Glozman:1997jy}   
in contrast to the one gluon exchange (OGE) model.
However, it cannot explain the hyperfine splitting in mesons, because
it does not explicitly contain a quark-antiquark interaction.

An SU(4) classification of pentaquarks and its decomposition into SU(3) submultiplets, by selecting 
those with the charm quantum number $C$ = 0, has been considered in Ref. \cite{Ortiz-Pacheco:2018ccl}
and several properties as  mass spectrum,  magnetic moments and photocouplings have been studied.
Other approaches can be found in the review papers  \cite{Chen:2016qju,Chen:2022asf}. 

Presently, the spin and parity of the narrow structures  $P^+_c(4312)$, $P^+_c(4440)$ and $P^+_c(4457)$ 
remains to be established experimentally.

In this context it is reasonable to expect  a future observation of the bottom analogues of the hidden charm
pentaquarks.  An optimistic note is that
the naming convention of a possible $uudb \bar b$ pentaquark has  already been listed in Ref. \cite{Gershon:2022xnn}.
as $P^N_{\Upsilon}(mass)$. 
Here we explore the spectrum of the  $uudb \bar b$ pentaquark in a constituent quark model with a
flavor-spin hyperfine interaction  \cite {Stancu:2019qga} which is now extended from SU(4) to SU(5)

Let us first recall a few  previous studies for hidden bottom pentaquarks.
In an early exploratory paper \cite{Wu:2010rv}
hidden bottom pentaquarks have been studied as hadronic $\Sigma_b B$ molecules and a few 
narrow resonances were found around 11 GeV.

Subsequently the 2015 LHCb observation of the $P_c!4380)$ and $P_c(4450)$ pentaquarks
stimulated interest in the study of hidden bottom pentaquarks. For example in Ref. \cite{Kopeliovich:2015vqa}
the mass of the hidden bottom $uud b \bar b$  pentaquark was estimated around 10.8 GeV in a 
simplified phenomenological model.
 
In Ref. \cite{Wu:2017weo} properties of hidden bottom pentaquarks $qqqb \bar b$ $(q = u,d,s)$ have been explored
in the framework of a simple quark model with chromomagnetic interaction.
which gave
satisfactory results for the 2015 LHCb $P_c$ resonances.
The theoretical masses were compared to the $\Upsilon p$, $\Upsilon \Delta$ and $\Sigma_b B$ thresholds.
A bound state with $I = 1/2$ and $J^P$ = $3/2^-$ was found. 

In Ref. \cite{Yang:2018oqd} a chiral quark model which successfully explains meson and baryon
phenomenology was applied to the hidden bottom sector to search for possible bound states 
with isospin 1/2 and 3/2. Several candidates were found for negative parity states.

In Ref. \cite{Yamaguchi:2017zmn} hidden bottom pentaquarks were investigated as
${\bar B}^{(*)} \Lambda_b$ and  ${\bar B}^{(*)} \Sigma_b$ molecules coupled to compact 5-quark states.
It was found that the pion exchange interaction is strong enough to produce resonant and bound states.

In Ref.  \cite {Stancu:2019qga} the extension to SU(4) 
has been made in the spirit of  Ref. \cite{Glozman:1995xy} where, in addition to the Goldstone bosons of 
the hidden approximate chiral symmetry of QCD,  the flavor exchange interaction was augmented  
by a phenomenological hyperfine flavor exchange of $D, D_s$ and $B$ mesons. 
The model provided a satisfactory description of heavy flavor baryons.

In this work  we consider the extension of the flavor-spin model \cite {Stancu:2019qga} from SU(4) to SU(5).
In addition to the Goldstone bosons of 
the hidden approximate chiral symmetry of QCD,  the flavor exchange interaction is augmented  
by an  exchange of open charm $D, D_s$, open bottom $B, B_s$ and 
open charm-bottom $B_c$ mesons. For consistency with the SU(5)  algebra the hidden-charm 
and the hidden-bottom mesons $\eta_c$ and $\eta_b$
were added too.

The paper is organized as follows. In Sec. \ref{Hamiltonian} we introduce the 
model Hamiltonian and  the two-body matrix elements of the FS interaction 
corresponding to SU(5) and needed for studying the $uud b \bar b$ pentaquark. 
Sec. \ref{coord} describes the orbital part of the 
four quark subsystem constructed to be translationally invariant both for positive 
and negative parity states.
Sections \ref{ke},\ref{linearcon} and \ref{FSint}
summarize analytic formulas.
Sec. \ref{numerical} contains numerical results
and a comparison with previous studies. 
The last section is devoted to
conclusions. 
Appendix 
\ref{Casimir}  is a reminder of useful group theory formulas for SU(n).
Appendix 
\ref{baryons} exhibits a variational solution for the baryon masses relevant for the 
present study.
Appendix \ref{lambda} reproduces the SU(5) generators $\lambda_i$ used in the calculation of the matrix elements 
of the flavor-spin interaction.

\section{The Hamiltonian}\label{Hamiltonian}
Here we closely follow the description of the model as presented in Ref. \cite{Stancu:2019qga}.
The  Hamiltonian has the general form \cite{Glozman:1995fu}
\begin{eqnarray}
H &=& \sum_i m_i + \sum_i\frac{{\vec p}_{i}^2}{2m_i} 
- \frac {(\sum_i {\vec p}_{i})^2}{2\sum_i m_i} + \sum_{i<j} V_{\text{conf}}(r_{ij}) \nonumber\\
&+& \sum_{i<j} V_\chi(r_{ij}),
\label{ham}
\end{eqnarray}
with $m_i$ and ${\vec p}_{i}$ 
denoting the quark masses and momenta respectively
and $r_{ij}$ the distance between the  quarks $i$ and $j$. 
The Hamiltonian contains the internal kinetic energy and the linear confining interaction 
\begin{equation}
 V_{\text{conf}}(r_{ij}) = -\frac{3}{8}\lambda_{i}^{c}\cdot\lambda_{j}^{c} \, C\, r_{ij} \, .
\label{conf}
\end{equation}
The SU(5) extension of the hyperfine interaction $V_\chi(r_{ij})$ has the following form
\begin{eqnarray}
V_\chi(r_{ij})
&=&
\left\{\sum_{F=1}^3 V_{\pi}(r_{ij}) \lambda_i^F \lambda_j^F \right. +  \sum_{F=4}^7 V_{K}(r_{ij}) \lambda_i^F \lambda_j^F 
\nonumber \\
&+& \left.  V_{\eta}(r_{ij}) \lambda_i^8 \lambda_j^8 
+V_{\eta^{\prime}}(r_{ij}) \lambda_i^0 \lambda_j^0\right. 
\nonumber \\
&+& \left. \sum_{F=9}^{12} V_{D}(r_{ij}) \lambda_i^F \lambda_j^F\right.     
+ \left. \sum_{F=13}^{14} V_{D_s}(r_{ij}) \lambda_i^F \lambda_j^F \right.
\nonumber \\
&+& \left. V_{\eta_c}(r_{ij}) \lambda_i^{15} \lambda_j^{15}\right.
\nonumber \\
&+& \left. \sum_{F=16}^{19} V_{B}(r_{ij}) \lambda_i^F \lambda_j^F\right. 
+ \left. \sum_{F=20}^{21} V_{B_s}(r_{ij}) \lambda_i^F \lambda_j^F\right.
+ \left. \sum_{F=22}^{23} V_{B_c}(r_{ij}) \lambda_i^F \lambda_j^F\right.
\nonumber \\
&+& \left. V_{\eta_b}(r_{ij}) \lambda_i^{24} \lambda_j^{24}  
\right\}
\vec\sigma_i\cdot\vec\sigma_j, 
\label{VCHI}
\end{eqnarray}
\noindent
with the SU(5) generators $\lambda^F_i$ ($F$ = 1,2,...,24)
and
$\lambda^0_i $ 
proportional to the unit matrix. 
In the SU(5) version
the interaction (\ref{VCHI})
contains $\gamma = \pi, K, \eta, D, D_s, \eta_c, B, B_s, B_c, \eta_b $ and $\eta '$
meson-exchange terms.  Every $V_{\gamma} (r_{ij})$ is
a sum of two contributions: a Yukawa-type potential containing
the mass $\mu_{\gamma}$ of the exchanged meson and a short-range contribution of opposite
sign, the role of which is crucial in baryon spectroscopy. 
For a given meson $\gamma$ the meson exchange potential is
\begin{eqnarray}\label{radialform}
V_\gamma (r) &=&
\frac{g_\gamma^2}{4\pi}\frac{1}{12m_i m_j}
\{\theta(r-r_0)\mu_\gamma^2\frac{e^{-\mu_\gamma r}}{ r} \nonumber\\
&-& \frac {4}{\sqrt {\pi}}
\alpha^3 \exp(-\alpha^2(r-r_0)^2)\}, 
\end{eqnarray}
where $\mu_\gamma$ is the mass of the exchanged meson, $m_i$  the quark mass,
$\frac{g_{\gamma}^2}{4\pi}$ the coupling constant and  
$r_0$ a phenomenological parameter defined in Ref. \cite{Glozman:1996wq}.

After integration in the flavor space the two-body matrix elements  
become products of the spin-spin operator and expressions depending on $V_\gamma (r)$.
Then, in the complete SU(5) extension containing five flavors, we have 
\begin{equation}\label{twobody}
V_{ij} =
{\vec {\sigma}}_i\ \cdot {\vec {\sigma}}_j\, 
\left\{ \renewcommand{\arraystretch}{2}
\begin{array}{cl}
 V^{uu}_{\pi} + \frac{1}{3} V^{uu}_{\eta} + \frac{1}{6} V^{uu}_{\eta_c} + \frac{1}{10} V^{uu}_{\eta_b},  &\hspace{0.3cm} \mbox{ $[2]_F, I = 1$} \\
2  V^{us}_K - \frac{2}{3} V^{us}_{\eta},
~ 2 V^{uc}_D - \frac{1}{2} V^{uc}_{\eta_c,},
~ 2 V^{ub}_B - \frac{2}{5} V^{ub}_{\eta_b} & \hspace{0.3cm} \mbox{ $[2]_F, I = \frac{1}{2}$} \\
2  V^{sc}_{D_s} - \frac{1}{2} V^{sc}_{\eta_c},
~ 2 V^{sb}_{B_s} - \frac{2}{5} V^{sb}_{\eta_b,},
~ 2 V^{cb}_{B_c} - \frac{2}{5} V^{cb}_{\eta_b} & \hspace{0.3cm} \mbox{ $[2]_F, I = 0$} \\
\frac{4}{3} V^{ss}_{\eta} + \frac{3}{2} V^{cc}_{\eta_c} + \frac{2}{5} V^{bb}_{\eta_b} &\hspace{0.3cm} \mbox{ $[2]_F, I = 0$} \\
 - 2  V^{sc}_{D_s} - \frac{1}{2} V^{sc}_{\eta_c},
~ - 2 V^{sb}_{B_s} - \frac{2}{5} V^{sb}_{\eta_b,},
~ - 2 V^{cb}_{B_c} - \frac{2}{5} V^{cb}_{\eta_b} & \hspace{0.3cm} \mbox{ $[2]_F, I = 0$} \\
 - 2  V^{us}_K - \frac{2}{3} V^{us}_{\eta},
~ - 2 V^{uc}_D - \frac{1}{2} V^{uc}_{\eta_c,},
~ - 2 V^{ub}_B - \frac{2}{5} V^{ub}_{\eta_b} &\hspace{0.3cm} \mbox{ $[11]_F, I = \frac{1}{2}$} \\
-3 V^{uu}_{\pi} + \frac{1}{3} V^{uu}_{\eta} + \frac{1}{6} V^{uu}_{\eta_c} + \frac{1}{10} V^{uu}_{\eta_b},  &\hspace{0.3cm} \mbox{ $[11]_F, I = 0$}
\end{array} \right.
\end{equation}
\noindent
where the quark pair $ij$ is either in a symmetric [2]$_F$ or in an antisymmetric [11]$_F$ flavor state
and the isospin $I$ is defined by the quark content. The upper index of 
$V$ exhibits the flavor of the
two quarks interchanging a meson specified by the lower index.
In every sum/difference of Eq. (\ref{twobody}) the upper index is the same
for all terms. Note that the $K, D, D_s, B, B_s$ and $B_c$ meson exchanges
contribute with a factor + 2 for symmetric pairs and  - 2  for antisymmetric pairs.

The present study is devoted to the hidden bottom pentaquark, the most expected to 
be searched for experimentally. 
This pentaquark contains $u, d$ and $b$ quarks
so that there are no $K$, $D$, $D_s$, $B_s$ and $B_c$ meson exchanges. 
Therefore only the terms of Eq. (\ref{twobody}) related to $\pi$, 
$\eta$, $B$, $\eta_c$ and $\eta_b$ exchanges will contribute
and in practice we ignore the contribution of  $\eta_c$, $\eta_b$ exchanges,
because little $u \bar u$ and $d \bar d$ are expected in real $\eta_c$, $\eta_b$. 
We recall that the scalar mesons   $\eta_c$ and $\eta_b$ used in the theoretical derivation 
of the expressions of Table  \ref{FOURQ}, based on the SU(5) algebra,
are defined in Appendix \ref{lambda}.

For hidden bottom pentaquarks 
we are left with a few parameters which
are chosen as follows.  For the light quarks we use the
parameters of Ref. \cite{Glozman:1996wq}  to which we add 
the $\mu_{B}$  mass and
the coupling constant $\frac{g_{Bq}^2}{4\pi}$. 
These are 
$$\frac{g_{\pi q}^2}{4\pi} = \frac{g_{\eta q}^2}{4\pi} =
\frac{g_{Bq}^2}{4\pi}=           0.67,\,\,
\frac{g_{\eta ' q}^2}{4\pi} = 1.206 , $$
$$r_0 = 0.43 \, \mbox{fm}, ~\alpha = 2.91 \, \mbox{fm}^{-1},~~
 C= 0.474 \, \mbox{fm}^{-2}, \, $$
$$ \mu_{\pi} = 139 \, \mbox{ MeV},~ \mu_{\eta} = 547 \,\mbox{ MeV},~
\mu_{\eta'} = 958 \, \mbox{ MeV},
 ~ \mu_{B} = 5279 \, \mbox{ MeV}.$$
The meson masses correspond to the experimental values from  the  Particle Data Group %
\cite{Workman:2022ynf}. 

The model of Ref. \cite{Glozman:1996wq} has previously been used to study the stability of
open flavor tetraquarks \cite{Pepin:1996id} and open flavor pentaquarks \cite{Genovese:1997tm,Stancu:1998sm}.
Accordingly,
for the quark masses we take the values determined variationally in 
Refs. \cite{Pepin:1996id,Genovese:1997tm} 
\begin{equation}\label{quarkmass}
 m_{u,d} = 340 \, \mbox{ MeV},~ m_b = 4660 \,\mbox{ MeV}.
\end{equation}
They were adjusted to satisfactorily reproduce the average mass ${\overline M} = (M + 3 M^*)/4$ = 5313 MeV of
the $B$ and $B^*$ mesons.


We can now present the contribution of the hyperfine interaction (\ref{VCHI})
to the pentaquark states which comes from the four-quark subsystem. 
The group theoretical structure of the states under consideration 
is specified in column 1 of Table \ref{FOURQ} for each state by the partitions
[F] for flavor, [S] for spin and  [FS] for flavor-spin.

In calculating the matrix elements of the hyperfine interaction (\ref{VCHI}) 
the first step is to decouple the flavor and spin parts of the wave functions of partitions $[f]_{FS}$.
This is done by using Clebsch-Gordan coefficients (isoscalar 
factors) of the permutation group $S_4$ \cite{Stancu:1999qr},
which allowed to reduce the  four-body to two-body matrix elements.

At this stage one needs the explicit form of the spin and flavor  wave functions
of the four-quark subsystem as 
specified in column 1 of Table \ref{FOURQ} for every  partition [FS]. 
The spin wave functions are trivial and not given in the paper.

The flavor wave functions can be obtained by analogy to the flavor wave functions 
$uudc$ which were derived in Appendix D of Ref. \cite{Stancu:2019qga}.
Indeed the similarity between Weyl's tableaux of the hidden charm and hidden bottom
four quark  subsystems allows the 
replacement of c by b in the flavor wave functions but this is just the starting point.
The corresponding wave functions have been written as linear combinations
of products of symmetric or antisymmetric two-quark pairs and used to calculate
the matrix elements of the hyperfine interaction using  
Eq. (5). This equation explicitly shows that the contribution  of a qc pair and a bq pair (q=u,d)
are entirely different from each other.
After lengthy calculations,
we have obtained 
the diagonal matrix elements of the flavor-spin  interaction (\ref{VCHI})
presented  in Table \ref{FOURQ}, every expression representing the contribution of
six pairs of quarks. All off-diagonal matrix elements vanish identically.

\begin{table}
\parbox{16cm}{\caption[matrix]{\label{FOURQ}
The hyperfine interaction $ V_{\chi}  $, Eq. (\ref{VCHI}), 
integrated in the flavor-spin space, for the quark subsystem $uudb$ with
$I$ = 1/2. 
${{V}^{q_a q_b}_{\gamma }}$ are defined in Eq.  (\ref{twobody})
where the upper index ${q_a q_b}$ indicates the flavor of the interacting quark pair.}}
%
%
\begin{tabular}{c|c}
\hline
  State &   $  V_{\chi}  $ \\[0.5ex]
\hline
$| 1 \rangle = |{{\left[{31}\right]}_{O}\
{\left[{22}\right]}_{F}{\left[{22}\right]}_{S}{\left[{4}\right]}_{FS}}\rangle$
  & \hspace{1mm} 15 $ V_{\pi} -  V^{uu}_{\eta} - 2 V^{uu}_{\eta'} - \frac{1}{2} V^{uu}_{\eta_c}
  -\frac{3}{10} V^{uu}_{\eta_b} + 12 V^{ub}_B  + \frac{6}{5} V^{ub}_{\eta_b} - 2  V^{ub}_{\eta'} $ 
\\[1.2ex]
$| 2 \rangle = |{{\left[{31}\right]}_{O}\
{\left[{31}\right]}_{F}{\left[{31}\right]}_{S}{\left[{4}\right]}_{FS}}\rangle$
 & \hspace{1mm} 3 $V_{\pi} + V^{uu}_{\eta} + 2 V^{uu}_{\eta'} + \frac{1}{2} V^{uu}_{\eta_c} + \frac{3}{10} V^{uu}_{\eta_b}
 + 14 V^{ub}_B  + 2  V^{ub}_{\eta_b}  - \frac{10}{3} V^{ub}_{\eta'}$ 
\\[1.2ex]
$| 3 \rangle =  |{{\left[{4}\right]}_{O}\
{\left[{211}\right]}_{F}{\left[{22}\right]}_{S}{\left[{31}\right]}_{FS}}\rangle$
  & \hspace{1mm}  $ 7 V_{\pi} - \frac{7}{9} V^{uu}_{\eta} - \frac{14}{9} V^{uu}_{\eta'}- \frac{7}{18} V^{uu}_{\eta_c}   
- \frac{7}{30} V^{uu}_{\eta_b}
+ \frac{22}{3} V^{ub}_B + \frac{22}{15} V^{ub}_{\eta_b} - \frac{22}{9} V^{ub}_{\eta'}$ 
\\[1.2ex]
$| 4 \rangle = |{{\left[{4}\right]}_{O}\
{\left[{31}\right]}_{F}{\left[{22}\right]}_{S}{\left[{31}\right]}_{FS}}\rangle$
  & \hspace{1mm}  $ -\frac{5}{3} V_{\pi} - \frac{5}{9} V^{uu}_{\eta} - \frac{10}{9} V^{uu}_{\eta'}  - \frac{5}{18} V^{uu}_{\eta_c}
  - \frac{5}{30} V^{uu}_{\eta_b} + \frac{22}{3}V^{ub}_B  + \frac{26}{15} V^{ub}_{\eta_b} - \frac{26}{9} V^{ub}_{\eta'}$
\\[0.9ex]
\hline
\end{tabular}
\end{table} 


It is useful to recall that
in the exact SU(5) limit, the flavor-spin interaction 
takes the following form  
\begin{equation}\label{schematic}
{V}_{\chi }\ =\ -\ {C}_{\chi }\ \sum\limits_{ i\ <\ j}^{} {\lambda }_{
i}^{ F} \cdot {\lambda }_{ j}^{ F} \ {\vec {\sigma }}_{i} \cdot {\vec {\sigma}}_{j},
\end{equation}
with $C_{\chi}$
an equal strength constant for all pairs.  
To reproduce the exact SU(5) limit of  the hyperfine interaction 
in Table \ref{FOURQ}
one has to take $V_{\pi}$ = $V^{uu}_{\eta}$ =  $V^{uu}_{\eta_c}$ =  $V^{uu}_{\eta_b}$
 = $V^{ub}_B$ = $V^{ub}_{\eta_b}$  =  - ${C}_{\chi }$
and $V^{uu}_{\eta'}$ = $V^{ub}_{\eta'}$  = 0. 
One obtains  -  $ \frac{132}{5} C_{\chi}$, -  $ \frac{104}{5} C_{\chi}$, 
 -  $ \frac{72}{5} C_{\chi}$ and   -  $ \frac{32}{5} C_{\chi}$     respectively.
These values can be checked with the formulas given in Appendix \ref{Casimir}.
They suggest that the lowest state is $|1 \rangle$. 
We shall see that
even with a broken SU(5) symmetry 
the lowest state of  $uudb \bar b$ has positive parity
because it acquires the largest
attraction due to the FS interaction, similar to  the 
$uudc \bar c$ hidden charm pentaquark  \cite{Stancu:2019qga} or the open charm
$uudd \bar c$ pentaquark \cite{Stancu:1998sm}.


\section{Orbital space}\label{coord}

As first introduced in Ref. \cite{Stancu:1998sm} 
the orbital wave functions are defined in terms of 
four internal Jacobi coordinates for pentaquarks chosen a
\begin{eqnarray}\label{coordin}
\begin{array}{c}\vec{x}\ =\ {\vec{r}}_{1}\ -\ {\vec{r}}_{2}\ ,\, \\ 
\vec{y}\ =\
{\left({{\vec{r}}_{1}\ +\ {\vec{r}}_{2}\ -\ 2{\vec{r}}_{3}}\right)/\sqrt
{3}} ,\, \\
\vec{z}\ =\ {\left({{\vec{r}}_{1}\ +\ {\vec{r}}_{2}\ +\ {\vec{r}}_{3}\ -\
3{\vec{r}}_{4}}\right)/\sqrt {6}} ,\, \\  
\vec{t}\ =\
{\left({{\vec{r}}_{1}\
+\ {\vec{r}}_{2}\ +\ {\vec{r}}_{3}+\ {\vec{r}}_{4}-\
4{\vec{r}}_{5}}\right)/\sqrt {10}} ,
\end{array}
\end{eqnarray}
where 1,2,3 and 4 are the quarks and 5 the antiquark so that
$t$ gives the distance between the antiquark and the center of mass coordinate 
of the four-quark subsystem. 


\subsection{ $P$ = + 1 pentaquarks}

The parity of the
pentaquark is given by P$\ = {\left({-}\right)}^{{\ell\ +\ 1}}$, where  $\ell$ is the orbital angular momentum.
The lowest positive parity pentaquark contains one unit of orbital excitation 
in the four-quark subsystem and has the  symmetry ${\left[{31}\right]}_{O}$. 
The method of
constructing translationally invariant states of definite permutation symmetry
containing a unit of angular momentum was first given in Ref. \cite{Stancu:1998sm}
and later revised in  Ref. \cite{Stancu:2019qga}.
The 
three independent states  denoted below by $\psi_i$, which define the 
basis vectors   of the irreducible representation 
${\left[{31}\right]}_O$   in 
terms of shell model states 
$ \left\langle{\vec{r}\left|{n \ell m}\right.}\right\rangle$ where $n$ = 0, $\ell$ = 1, 
are
\begin{eqnarray}
{\psi }_{1} = 
\left\langle{\vec{x}\left|{000}\right.}\right
\rangle\left\langle{\vec{y}\left|{000}\right.}\right\rangle\left
\langle{\vec{z}\left|{010}\right.}\right\rangle,
\end{eqnarray}
\begin{eqnarray}
{\psi }_{2} = 
\renewcommand{\arraystretch}{0.5}
\left\langle{\vec{x}\left|{000}\right.}\right
\rangle\left\langle{\vec{y}\left|{010}\right.}\right\rangle\left
\langle{\vec{z}\left|{000}\right.}\right\rangle,
\end{eqnarray}
\begin{eqnarray}
{\psi }_{3} = 
\left\langle{\vec{x}\left|{010}\right.}\right
\rangle\left\langle{\vec{y}\left|{000}\right.}\right\rangle\left
\langle{\vec{z}\left|{000}\right.}\right\rangle.
\end{eqnarray}

\noindent 
The  pentaquark orbital wave functions $\psi_i^5$ are  obtained by multiplying
each $\psi_i$ from above by the wave function
$\left\langle{\vec{t}\left|{000}\right.}\right\rangle$ which describes the
relative motion between the four-quark subsystem and the antiquark $\overline{c}$.
Assuming an exponential behavior we introduce two variational parameters, 
$a$ for the internal motion of the four-quark subsystem  and
$b$ for the relative motion between the subsystem $qqqc$ and $\overline{c}$. We explicitly have
\begin{equation}\label{psi1}
{\psi }_{1}^{5}\ =\ N\ \exp\ \left[{-\ {\frac{a}{2}}\ \left({{x}^{2}\ +\
{y}^{2}\ +\ {z}^{2}}\right)\ -\ {\frac{b}{2}}\ {t}^{2}}\right]\ z\ {Y}_{10}\
\left({\hat{z}}\right),
\end{equation}
\begin{equation}\label{psi2}
{\psi }_{2}^{5}\ =\ N\ \exp\ \left[{-\ {\frac{a}{2}}\ \left({{x}^{2}\ +\
{y}^{2}\ +\ {z}^{2}}\right)\ -\ {\frac{b}{2}}\ {t}^{2}}\right]\ y\ {Y}_{10}\
\left({\hat{y}}\right),
\end{equation}
\begin{equation}\label{psi3}
{\psi }_{3}^{5}\ =\ N\ \exp\ \left[{-\ {\frac{a}{2}}\ \left({{x}^{2}\ +\
{y}^{2}\ +\ {z}^{2}}\right)\ -\ {\frac{b}{2}}\ {t}^{2}}\right]\ x\ {Y}_{10}\
\left({\hat{x}}\right),
\end{equation}
where
\begin{equation}
N\ =\ {\frac{{2}^{3/2}{a}^{11/4}{b}^{3/4}}{{3}^{1/2}{\pi }^{5/2}}}.
\end{equation}

\subsection{ $P$ = - 1 pentaquarks}

The orbital wave function of 
the lowest negative parity state described by the
$s^4$ configuration of symmetry $[4]_O$  is defined as 
\begin{equation}\label{phi}
\phi_0 = \ N_0\ \exp\ \left[{-\ {\frac{a}{2}}\ \left({{x}^{2}\ +\
{y}^{2}\ +\ {z}^{2}}\right)\ -\ {\frac{b}{2}}\ {t}^{2}}\right],
\end{equation}
with
\begin{equation}\label{phinorm}
N_0\ =\ ({\frac{a}{\pi}})^{9/4} (\frac{b}{\pi})^{3/4}.
\end{equation}


\section{Kinetic energy}\label{ke}

The kinetic energy  $T$  of the Hamiltonian (\ref{ham})
can be calculated analytically.

For {\it $P$} = +1 states 
the  expectation value of the kinetic energy is defined by the average over the three wave functions
defined by Eqs. (\ref{psi1})-(\ref{psi3}). One obtains
\begin{eqnarray}\label{TA}
\begin{array}{lcl}\left\langle{T}\right\rangle\ &=&\ {\frac{1}{3}}\
\left[{\left\langle{{\psi }_{1}^{5}\left|{T}\right|{\psi\
}_{1}^{5}}\right\rangle\ +\ \left\langle{{\psi }_{2}^{5}\left|{T}\right|{\psi
}_{2}^{5}}\right\rangle\ +\ \left\langle{{\psi }_{3}^{5}\left|{T}\right|{\psi
}_{3}^{5}}\right\rangle}\right]\\
\\ &=&\ {\hbar }^{2}\ \left({{\frac{11}{2{\mu }_{1}}}\ a\ +\ {\frac{3}{{2\mu
}_{2}}}\ b}\right),
\end{array}
\end{eqnarray}
with
\begin{eqnarray}\label{reduced}
{\frac{4}{{\mu }_{1}}}\
 =  \ \frac{3}{{m}_{q}}\  + \ \frac{1}{{m}_{Q}},\ 
\end{eqnarray}
and  
\begin{equation}\label{redmass}
{\frac{5}{{\mu }_{2}}}\ =\ {\frac{1}{{\mu }_{1}}}\ +\ {\frac{4}{{m}_{Q}}},
\end{equation}
where $q = u, d$ and $Q = b$ 
with masses defined by Eq. (\ref{quarkmass}). 


For {\it P} = - 1 states there is no orbital excitation and the orbital wave function 
of the four-quark subsystem has the permutation symmetry  $[4]_O$.
In this case  Eq. (\ref{phi}) gives
\begin{eqnarray}\label{TC}
\left\langle{T}\right\rangle\ = \ {\hbar }^{2}\ \left({{\frac{9}{2{\mu }_{1}}}\ a\ +\ {\frac{3}{{2\mu
}_{2}}}\ b}\right),
\end{eqnarray}
with $\mu_{1}$ and $\mu_{2}$  as above.

\section{Confinement}\label{linearcon}

By integrating in the color space, the expectation value of the confinement interaction (\ref{conf}) 
has the same form as that of the $uudc \bar c $ system  \cite{Stancu:2019qga}
\begin{equation}\label{confin}
\left\langle{{V}_{conf}}\right\rangle\ =\ {\frac{C}{2}}\ \left({6\
\left\langle{{r}_{12}}\right\rangle\ +\ 4\
\left\langle{{r}_{45}}\right\rangle}\right)
\end{equation}
where $\langle{{r}_{ij}}\rangle$ is the interquark distance and
the coefficients 6 and 4 account for the number of quark-quark and
quark-antiquark pairs, respectively, 
but the expression for $\langle{{r}_{ij}}\rangle$ depends on parity.

For {\it P} = + 1 one has
\begin{equation}\label{r12}
\left\langle{{r}_{ij}}\right\rangle\ =\ {\frac{1}{3}}\
\left[{\left\langle{{\psi }_{1}^{5}\left|{{r}_{ij}}\right|{\psi
}_{1}^{5}}\right\rangle\ +\ \left\langle{{\psi
}_{2}^{5}\left|{{r}_{ij}}\right|{\psi }_{2}^{5}}\right\rangle\ +\
\left\langle{{\psi }_{3}^{5}\left|{{r}_{ij}}\right|{\psi
}_{3}^{5}}\right\rangle}\right],
\end{equation}
where $i,j$ = 1,2,3,4,5 ($ i \neq j$).
An analytic evaluation gives
\begin{equation}
\left\langle{{r}_{12}}\right\rangle\ =\ {\frac{20}{9}}\ \sqrt {{\frac{1}{\pi
 a}}},
\end{equation}
and 
\begin{equation}
\left\langle{{r}_{45}}\right\rangle\ =\ {\frac{1}{3\sqrt {2\pi }}}\
\left[{2\sqrt {{\frac{3}{a}}\ +\ {\frac{5}{b}}}\ +\ \sqrt {5b}\
\left({{\frac{1}{2a}}\ +\ {\frac{1}{b}}}\right)}\right].
\end{equation}

For {\it P} = - 1 there is no orbital excitation so that 
the four quarks are in the
$s^4$ configuration of permutation symmetry $[4]$.
The expectation value of the confinement interaction is given by Eq. (\ref{confin})
as well, with 
\begin{equation}\label{r12ground}
\left\langle{{r}_{12}}\right\rangle\ = \sqrt {{\frac{4}{\pi a}}},
\end{equation}
and 
\begin{equation}\label{r45ground}
\left\langle{{r}_{45}}\right\rangle\ =\ {\frac{1}{\sqrt {2\pi }}}\
\sqrt{{\frac{3}{a}}\ +\ {\frac{5}{b}}}.
\end{equation}


\section{Flavor-spin interaction}\label{FSint}

For integrating the expressions  of  Table \ref{FOURQ} 
in the orbital space one has to decouple 
the orbital part of the wave function $[f]_O$ from the part containing the other degrees of
freedom by  again using Clebsch-Gordan coefficients (isoscalar factors) of the permutation group $S_4$ \cite{Stancu:1999qr}
but this time related to the orbital and the flavor-spin space. The use of the permutation properties
of the translationally invariant orbital wave functions is necessary at this stage.

Including the orbital space, it turns out that
for positive parity states   with one unit of orbital excitation  
the result is a linear combination of  two-body matrix elements of type
$\left\langle{ss\left|{{V}^{q_a q_b}_{\gamma }}\right|ss}\right\rangle\ ,\
\left\langle{sp\left|{{V}^{q_a q_b}_{\gamma }}\right|sp}\right\rangle$ and
$\left\langle{sp\left|{{V}^{q_a q_b}_{\gamma }}\right|ps}\right\rangle$.
For negative parity states the hyperfine interaction contains only two-body matrix elements of type
$\left\langle{ss\left|{{V}^{q_a q_b}_{\gamma }}\right|ss}\right\rangle$.
In every term  $q_a q_b$
is a pair of quarks from Eq. (\ref{twobody}).
 
\section{Results and discussion}\label{numerical}

The lowest part of the mass spectrum of 
the $uudb \bar b$ pentaquark given by Hamiltonian of Sec. \ref{Hamiltonian}
has been calculated variationally  
with the wave functions 
described in Sec. \ref{coord}, containing the parameters $a$ and $b$.
The flavor-spin part of
each wave function is a  product of a four quarks subsystem state 
defined in Table \ref{FOURQ}  and the
bottom antiquark wave function denoted by $|\bar b \rangle$. 
The total angular momentum is  $\vec{J}\ =\ \vec{L}\ +\ \vec{S}\ +\
{\vec{s}}_{Q}$, with $\vec{L}$ and $\vec{S}$ the angular momentum and the spin of the
four-quark cluster respectively and $\vec{s}_Q$ is the spin of the heavy antiquark.
In numerical calculations the expressions of the hyperfine interaction of Table \ref{FOURQ}
are simplified.
The contribution of  $V^{uu}_{\eta_c}$, $V^{uu}_{\eta_b}$ and $V^{ub}_{\eta_b}$ are neglected 
because little
$u \bar u$ and $d \bar d$ are expected in  ${\eta_c}$ and ${\eta_b}$.   We have also neglected 
$V^{ub}_{\eta'}$  assuming a little $b \bar b$ component in $\eta'$. 
Thus, in the expressions of Table \ref{FOURQ} we took 
\begin{equation}
V^{uu}_{\eta_c} = V^{uu}_{\eta_b} =  V^{ub}_{\eta_b} = V^{ub}_{\eta'} =  0.
\end{equation}

The numerical results are presented in Table \ref{resultA}.  
The eigenvalues of  $|1 \rangle |\bar b \rangle $
and    $|2 \rangle |\bar b \rangle $  states are degenerate for the allowed values of $J$ in each case.  
For  $|2 \rangle |\bar c \rangle $ the states with $J^P =  {1/2}^+$ and  ${3/2}^+$
have multiplicity 2. One can see that the lowest state has positive parity like for  
 $uud c \bar c$ \cite{Stancu:2019qga} an also for the $uudd \bar c$ pentaquark studied 
long time ago \cite{Stancu:1998sm}.

The optimal value of the parameter  $a$ varies with the state but $b$ is the same for all states.
The parameter a gives a measure of the compactness of the four quark subsystem because the 
mean distance between the quarks  is inverse proportional to the square root of $a$, see Eq. (\ref{r12ground}).
Thus the excited state $|4 \rangle$ is less compact than the others. 

\begin{table*}
\parbox{16cm}{\caption[matrix]{\label{resultA}
Lowest positive and negative parity  $uud b \bar b$ pentaquarks
of isospin $I$ = 1/2  and symmetry structures $|1 \rangle$, $|2 \rangle$, $|3 \rangle$
and $|4 \rangle$ defined in Table \ref{FOURQ}.
Column 1 gives the state,  column 2  the parity and total angular momentum, 
column 3 and 4 the
optimal variational parameters associated to the wave functions defined in Sec. \ref{coord},
 column 5 the calculated mass and columns 6 and 7 the relative distance between quarks/antiquarks.}}
\begin{tabular}{cccccccc}
\hline
State & \hspace{3mm}  $J^P$ & \multicolumn{2}{c}{Variational parameters} & \hspace{2mm}
Mass  & \hspace{4mm} $\langle r_{12} \rangle$ &  \hspace{3mm} $\langle r_{45} \rangle$\\
 & \hspace{6mm} & \hspace{2mm} a (fm$^{-2}$) & b (fm$^{-2}$) &  \hspace{2mm} (MeV)&   \hspace{3mm} (fm) &  \hspace{2mm} (fm)\\
\hline 
$ | 1 \rangle ~|\overline b \rangle$  & \hspace{2mm}  $\frac{1}{2}^+$, $\frac{3}{2}^+$ 
& \hspace{6mm} 1.900 & 1.387 & \hspace{2mm} 10961 & \hspace{4mm} 0.910  & \hspace{4mm} 0.950 \\[1.1ex]
$ | 2 \rangle ~|\overline b\rangle$  
& \hspace{2mm}  $\frac{1}{2}^+$, $\frac{3}{2}^+$, $\frac{5}{2}^+$ 
& \hspace{6mm} 1.387 & 1.387 & \hspace{2mm} 11133 & \hspace{4mm} 1.065 & \hspace{4mm} 1.018  \\[1.1ex]
$ | 3 \rangle ~|\overline b \rangle$  & \hspace{2mm}  $\frac{1}{2}^-$ 
& \hspace{6mm} 1.027 & 1.387 & \hspace{2mm} 11112 & \hspace{4mm} 1.113 & \hspace{4mm} 1.019   \\[1.1ex]
$ | 4 \rangle ~|\overline b \rangle$  & \hspace{2mm}  $\frac{1}{2}^-$
& \hspace{6mm} 0.514 & 1.387 & \hspace{2mm} 11334  & \hspace{4mm} 1.575 & \hspace{4mm} 1.226 \\[1.1ex]
\hline
\end{tabular}
\end{table*}

The detailed contribution of different parts of the Hamiltonian are given in Table \ref{parts}.
This shows that,
although the kinetic energy of the lowest positive parity state named $ | 1 \rangle ~|\overline b \rangle$
is about twice larger than that of the lowest negative parity state named  $ | 3 \rangle ~|\overline b \rangle$,  
the flavor-spin interaction overcomes this excess and generates a lower 
eigenvalue of 10961 MeV for the   ${\left[{31}\right]}_{O}$  state  with an  $s^3p$ configuration  than for 
${\left[{4}\right]}_{O}$  with an $s^4$ configuration, the eigenvalue of which becomes 11112 MeV.
Another interesting remark is that the contribution of $V^{\chi}$ to the mass of the
higher negative parity state $ | 4 \rangle ~|\overline b \rangle$ is repulsive. 
This is due to the fact that the dominant pion exchange in the expression 
of  Table \ref{FOURQ} corresponding to  $ | 4 \rangle $ has negative sign, contrary 
to the other three cases. 

In Table \ref{resultA} we also exhibited the distance $\langle r_{12} \rangle$ between a pair of quarks
and the distance $\langle r_{45} \rangle$ between a quark and the antiquark. For all considered states 
they have  small comparable values which indicates that the pentaquark is rather compact.
For the state $ | 4 \rangle ~|\overline b \rangle$  the distance $\langle r_{12} \rangle$ is larger
which can be explained by  the larger contribution of $V_{conf}$.


\begin{table}
\caption{\label{parts} Partial contributions of the Hamiltonian of Sec. \ref{Hamiltonian}
to the calculated masses of the pentaquarks given in Table \ref{resultA}.}
\begin{tabular}{cccccc}
\hline
State   & \hspace{1mm}   Parity &\hspace{1mm} $ K.E. $     & \hspace{1mm}  $ V_{conf} $ & \hspace{2mm} $ V^{\chi} $\\
\hline
$ |1 \rangle |\overline b \rangle $ &  \hspace{2mm} + 1 & \hspace{2mm} 970   & \hspace{2mm}  432 
&  \hspace{2mm} - 781  \\[1.1ex]
$ |2 \rangle |\overline b \rangle $ &  \hspace{2mm} + 1 & \hspace{2mm} 723   & \hspace{2mm}  487 
&  \hspace{2mm} - 416  \\[1.1ex]
$ |3 \rangle |\overline b \rangle $ &  \hspace{2mm} - 1 & \hspace{2mm} 457   & \hspace{2mm}  501 
&  \hspace{2mm} - 186  \\[1.1ex] 
$ |4 \rangle |\overline b \rangle $ &  \hspace{2mm} - 1 & \hspace{2mm} 254  &  \hspace{2mm}  669 
&  \hspace{6mm} 71  \\[1.1ex]
\hline
\end{tabular}
\end{table}

Besides the level ordering it would be useful to present a few 
more general features of the hidden bottom pentaquark spectrum and compare it
with the spectrum of hidden charm pentaquark studied  in Ref. \cite{Stancu:2019qga}
based on the SU(4) flavor-spin model. In Table \ref{comparison}
we recall the masses of the three common states calculated for pentaquarks
with charm and bottom. It is relevant to look at the relative positions in 
a given spectrum, the absolute value being dependent on the quark input 
masses, as it it known. One can notice that the spectrum of the 
hidden bottom pentaquark is more compressed, in the sense that the relative
position of the levels is smaller than the relative positions in the hidden charm spectrum. 
In particular the mass difference  
$ |2 \rangle |\overline b \rangle $ - $ |1 \rangle |\overline b \rangle $
is  180 MeV for hidden charm and 172 MeV for hidden bottom pentaquarks. 
This looks natural because the b quark is heavier than the c quark . 

As the number of observed heavy flavor baryons is increasing there is hope that 
more heavy flavor pentaquarks will be searched for. In the present model it will be useful
to study the strange hidden-bottom pentaquark $udsb \bar b$ as well or pentaquarks with
two distinct heavy flavors as $udcb \bar b$, etc. to look for their stability
against strong decays and aspects of their spectra.

\begin{table}
\caption{\label{comparison} 
Pentaquark masses (MeV) denoted by M($uud Q \bar Q$)  ($Q = c, b$) having  quark flavor structures
defined in Table \ref{FOURQ}.
Column 1 gives the state, column 2 the spin and parity, column 3 from Ref. \cite{Stancu:2019qga}
and 4 the present masses. }
\begin{tabular}{cccccc}
\hline
State   & \hspace{1mm}   $J^P$  &\hspace{1mm}  M($uud c \bar c$)      & \hspace{1mm}   M($uud b \bar b$)  
\\
\hline
$ |1 \rangle |\overline Q \rangle $ &  \hspace{2mm} $\frac{1}{2}^+$, $\frac{3}{2}^+$ & \hspace{2mm} 4273   & \hspace{2mm}  10961 
\\[1.1ex]
$ |2 \rangle |\overline Q \rangle $ &  \hspace{2mm} $\frac{1}{2}^+$, $\frac{3}{2}^+$, $\frac{5}{2}^+$ & \hspace{2mm} 4453   & \hspace{2mm}  11133 
\\[1.1ex]
$ |3 \rangle |\overline Q \rangle $ &  \hspace{2mm}  $\frac{1}{2}^-$& \hspace{2mm} 4487   & \hspace{2mm}  11112 
\\[1.1ex] 
\hline
\end{tabular}
\end{table}


\section{Conclusions}

The present work is a natural extension  of that of Ref. \cite{Stancu:2019qga}
where the spectrum of the $uudc \bar c$ pentaquark has been analyzed using the SU(4) flavor-spin model.
Here we have studied the hidden bottom 
$uud b \bar b $  pentaquark spectrum  in  the SU(5) version of the flavor-spin model.
The model provides an isospin dependence and an internal structure of pentaquarks
contrary to the molecular scenario.  
For positive parity states the angular 
momentum is located in the internal motion of the four-quark subsystem
and it turns out that the lowest state has positive parity, 
as in the case of the hidden charm pentaquark. 
In particular, we found that the coupling between the two negative parity states $ |3 \rangle |\overline b \rangle $
and $ |4 \rangle |\overline b \rangle $ 
vanish identically although the spin part is the same.
The lowest state has a mass of 10961 MeV and is located below the experimental thresholds 
 $\Sigma_b + B$  (11093 MeV) and  $ \Sigma^*_b + B $ (11112 MeV), thus it is stable  
against the corresponding strong decays.   
It can decay into $ \Lambda_b + B^*$ (10944 MeV), $\Lambda_b + B $ (10899 MeV), $\Upsilon + p$ (10398 MeV)
or $\eta_b + p$ (10337 MeV).   
The lowest state is  stable as well  against the theoretical threshold   $ \Sigma_b + B$.
We recall that the mass of the quark $b$, given in Eq. (\ref{quarkmass}),  has
been obtained from fitting the calculated mass of the $B$ and $B^*$ mesons which are degenerate,
to the average  experimental
mass 
which is 5312 MeV. Together with the calculated mass of $\Sigma_b$
(Appendix \ref{baryon}) one obtains 11049 MeV, i.e. above the lowest state mass. 
The lowest mass obtained in our study is in a range comparable to previous studies
  \cite{Wu:2010rv,Kopeliovich:2015vqa,Wu:2017weo,Yang:2018oqd,Yamaguchi:2017zmn}. The main difference 
is that the lowest state has positive parity in the present case.
Possible future observations of $uud b \bar b$ pentaquarks will be essential
to discriminate  between the existing interpretations or
inspire new developments.

We hope that in the future the present work could stimulate some interest in studying pentaquarks
containing quarks/antiquarks  of five distnct flavors in other models.


\appendix


\section{Exact SU(5) limit}\label{Casimir}

The exact SU(5) limit is useful in checking the integration in the flavor space,
made in Table  \ref{FOURQ}.  
In this limit every expectation value of  Table \ref{FOURQ}  reduces to the expectation value of Eq.  (\ref{schematic})
and one can use the following
formula \cite{Ortiz-Pacheco:2018ccl}
\begin{equation}\label{lambdasigma}
\langle ~\sum_{i<j} \lambda^F_i \cdot \lambda^F_j \vec{\sigma}_i \cdot \vec{\sigma}_j~\rangle = 4
C_{2}^{SU(2n)} - 2 C_{2}^{SU(n)} -
\frac{4}{n} C_{2}^{SU(2)} - k \frac{3(n^2 - 1)}{n}
\end{equation}
where $n$ is the number flavors and  $k$ the number of quarks, here $n$ = 5 and $k$ = 4.  $C_{2}^{SU(n)}$ is the
Casimir operator eigenvalues of $SU(n)$ which can be derived from the
expression  \cite{Stancu:1997dq} :
\begin{eqnarray}\label{casimiroperator}
C_{2}^{SU(n)} = \frac{1}{2} [f_1'(f_1'+n-1) + f_2'(f_2'+n-3) + f_3'(f_3'+n-5)
 \nonumber \\
+f_4'(f_4'+n-7) + ... + f_{n-1}'(f_{n-1}'-n+3) ] - \frac{1}{2n}
(\sum_{i=1}^{n-1} f_i')^2
\label{casimir}
\end{eqnarray}
where $f_i'= f_i-f_n$, for an irreducible representation given by the
partition $[f_1,f_2,...,f_n]$.
Eq. (\ref{lambdasigma}) has been previously used for $n$ = 3 and $k$ = 6 in Ref.  \cite{Stancu:1997dq}. 


\section{The baryons}\label{baryons}

\begin{table*}
\caption{\label{baryon} Masses of ground state baryons with the
flavor-spin interaction of Sec. \ref{Hamiltonian}.
Column 1 gives the baryon, column 2 the isospin, column 3 the spin and parity 
column 4 the calculated mass, column 5 the variational parameter 
and the last column the experimental mass.}
\begin{tabular}{ccccccc}
\hline
Baryon & \hspace{2mm}$I$ & $J^P$ & {Calc. Mass (MeV)}     & \hspace{1mm} a(fm$^{-2}$) & \hspace{0mm} Exp.mass (MeV)\\
\hline
$ \Lambda_b $ & \hspace{1mm} $ 0 $ & \hspace{2mm}  $\frac{1}{2}^+$ & \hspace{5mm} 5585   & \hspace{2mm}  2.080 
&   5620  \\[1.1ex]
$ \Sigma_b $ & \hspace{1mm} $ 0 $ & \hspace{2mm}  $\frac{1}{2}^+$ & \hspace{5mm} 5747  & \hspace{2mm}  1.284 
&   5813  \\[1.1ex]
$ \Sigma^*_b     $ & \hspace{1mm} $ 0 $ & \hspace{2mm}  $\frac{3}{2}^+$ & \hspace{5mm} 5773   & \hspace{2mm}  1.284 
&   5832   \\[1.1ex] 
\hline
\end{tabular}
\end{table*}

The masses of $\Lambda_b$, $\Sigma_b$ and $\Sigma_b ^*$ relevant for this study 
were calculated variationally with a radial wave function of the form 
$\phi \propto$   exp$[-\frac{a}{2}(x^2 + y^2)]$
with the variational parameter  $a$
and the coordinates $x$ and $y$ defined by Eq. (\ref{coordin}).
We took $V^{uu}_{\eta_c} =   V^{uu}_{\eta_b} =  V^{ub}_{\eta_b} =  V^{ub}_{\eta'} = 0$
like for pentaquarks. 
The results are indicated in Table \ref{baryon} together with the experimental
masses. These masses can be used to estimate theoretical thresholds consistent with the model,
as it is done in the conclusions. 
One can see that the experimental $ \Sigma^*_b$ - $ \Sigma_b$ splitting is better reproduced
than that obtained from the phenomenological flavor-spin interaction of Ref.  \cite{Glozman:1995xy},
where a $B$-meson exchange was simply assumed by analogy to $D$-meson exchange.


\section{SU(5) generators}\label{lambda}

Here we reproduce the $\lambda_i$ matrices which are the 
SU(5) generators in the fundamental representation of SU(5). 
Implementing them in Eq. (\ref{VCHI}) one can obtain the two-body matrix elements
of Eq. (\ref{twobody}) for each pair of quarks of a given flavor.

The first 15 matrices are an extension of the SU(4) generators \cite{Stancu:1991rc}
with one 0-row and one 0-column added.
We  have 
\begin{eqnarray}
\lambda_1   &  = & 
\left( 
\begin{array}{ccccc}
0 & 1 & 0 & 0 & 0\\
1 & 0 & 0 & 0 & 0\\ 
0 & 0 & 0 & 0 & 0 \\
0 & 0 & 0 & 0 & 0 \\
0 & 0 & 0 & 0 & 0 \end{array} 
\right)
,\,\,\,
\lambda_2 = 
\left(
\begin{array}{ccccc}
0 & - i & 0 & 0 & 0 \\
i & 0 & 0 & 0 & 0 \\ 
0 & 0 & 0 & 0 & 0 \\
0 & 0 & 0 & 0 & 0 \\
0 & 0 & 0 & 0 & 0 \end{array}
\right)
,\,\,\,
\lambda_3   =  
\left(
\begin{array}{ccccc}
1 & 0 & 0 & 0 & 0 \\
0 & - 1 & 0 & 0 & 0 \\ 
0 & 0 & 0 & 0 & 0 \\
0 & 0 & 0 & 0 & 0 \\
0 & 0 & 0 & 0 & 0 \end{array}
\right),
\end{eqnarray} 
\begin{eqnarray}
\lambda_4 = 
\left(
\begin{array}{ccccc}
0 & 0 & 1 & 0 & 0 \\
0 & 0 & 0 & 0 & 0 \\ 
1 & 0 & 0 & 0 & 0 \\
0 & 0 & 0 & 0 & 0 \\
0 & 0 & 0 & 0 & 0 \end{array}
\right)
,\,\,\,
\lambda_5 =  
\left(
\begin{array}{ccccc}
0 & 0 & - i & 0 & 0 \\
0 & 0 & 0 & 0 & 0 \\ 
i & 0 & 0 & 0 & 0 \\
0 & 0 & 0 & 0 & 0 \\
0 & 0 & 0 & 0 & 0 \end{array}
\right) 
,\,\,\,
\lambda_6 = 
\left(
\begin{array}{ccccc}
0 & 0 & 0 & 0 & 0 \\
0 & 0 & 1 & 0 & 0 \\ 
0 & 1 & 0 & 0 & 0 \\
0 & 0 & 0 & 0 & 0 \\
0 & 0 & 0 & 0 & 0 \end{array}
\right), 
\end{eqnarray}
\begin{eqnarray}
\lambda_7 = 
\left(
\begin{array}{ccccc}
0 & 0 & 0 & 0 & 0 \\
0 & 0 & - i & 0 & 0 \\ 
0 & i & 0 & 0 & 0 \\
0 & 0 & 0 & 0 & 0 \\
0 & 0 & 0 & 0 & 0 \end{array}
\right) 
,\,\,\,
\lambda_8    =
\frac{1}{\sqrt{3}}
\left(
\begin{array}{ccccc}
1 & 0 & 0 & 0 & 0 \\
0 & 1 & 0 & 0 & 0 \\ 
0 & 0 & - 2 & 0 & 0 \\
0 & 0 & 0 & 0 & 0 \\ 
0 & 0 & 0 & 0 & 0 \end{array}
\right) 
,\,\,\,
\lambda_9 = 
\left(
\begin{array}{ccccc}
0 & 0 & 0 & 1 & 0 \\
0 & 0 & 0 & 0 & 0 \\ 
0 & 0 & 0 & 0 & 0 \\
1 & 0 & 0 & 0 & 0 \\
0 & 0 & 0 & 0 & 0 \end{array}
\right), 
\end{eqnarray}
\begin{eqnarray}
\lambda_{10}    = 
\left(
\begin{array}{ccccc}
0 & 0 & 0 & - i & 0\\
0 & 0 & 0 & 0 & 0 \\ 
0 & 0 & 0 & 0 & 0 \\
i & 0 & 0 & 0 & 0 \\
0 & 0 & 0 & 0 & 0 \end{array}
\right) 
,\,\,\,
\lambda_{11} = 
\left(
\begin{array}{ccccc}
0 & 0 & 0 & 0 & 0 \\
0 & 0 & 0 & 1 & 0\\ 
0 & 0 & 0 & 0 & 0 \\
0 & 1 & 0 & 0 & 0 \\
0 & 0 & 0 & 0 & 0 \end{array}
\right) 
,\,\,\,
\lambda_{12} = 
\left(
\begin{array}{ccccc}
0 & 0 & 0 & 0 & 0 \\
0 & 0 & 0 & - i & 0 \\ 
0 & 0 & 0 & 0 & 0 \\
0 & i & 0 & 0 & 0 \\
0 & 0 & 0 & 0 & 0 \end{array}
\right), 
\end{eqnarray}
\begin{eqnarray}
\lambda_{13}  &  = & 
\left(
\begin{array}{ccccc}
0 & 0 & 0 & 0 & 0 \\
0 & 0 & 0 & 0 & 0 \\ 
0 & 0 & 0 & 1 & 0 \\
0 & 0 & 1 & 0 & 0 \\
0 & 0 & 0 & 0 & 0 \end{array}
\right) 
,\,\,\,
\lambda_{14} = 
\left(
\begin{array}{ccccc}
0 & 0 & 0 & 0 & 0 \\
0 & 0 & 0 & 0 & 0 \\ 
0 & 0 & 0 & - i & 0 \\
0 & 0 & i & 0 & 0 \\
0 & 0 & 0 & 0 & 0 \end{array}
\right) 
,\,\,\,
\lambda_{15} = 
\frac{1}{\sqrt{6}}
\left(
\begin{array}{ccccc}
1 & 0 & 0 & 0 & 0 \\
0 & 1 & 0 & 0 & 0 \\ 
0 & 0 & 1 & 0 & 0 \\
0 & 0 & 0 & - 3 & 0 \\
0 & 0 & 0 & 0 & 0 \end{array}
\right). 
\end{eqnarray}
The additional matrices are
\begin{eqnarray}
\lambda_{16} = 
\left(
\begin{array}{ccccc}
0 & 0 & 0 & 0 & 1 \\
0 & 0 & 0 & 0 & 0 \\ 
0 & 0 & 0 & 0 & 0 \\
0 & 0 & 0 & 0 & 0 \\
1 & 0 & 0 & 0 & 0 \end{array}
\right) 
,\,\,\,
\lambda_{17} = 
\left(
\begin{array}{ccccc}
0 & 0 & 0 & 0 & - i \\
0 & 0 & 0 & 0 & 0 \\ 
0 & 0 & 0 & 0 & 0 \\
0 & 0 & 0 & 0 & 0 \\
i & 0 & 0 & 0 & 0 \end{array}
\right) 
\lambda_{18} = 
\left(
\begin{array}{ccccc}
0 & 0 & 0 & 0 & 0 \\
0 & 0 & 0 & 0 & 1 \\ 
0 & 0 & 0 & 0 & 0 \\
0 & 0 & 0 & 0 & 0 \\
0 & 1 & 0 & 0 & 0 \end{array}
\right), 
\end{eqnarray}
\begin{eqnarray}
\lambda_{19} = 
\left(
\begin{array}{ccccc}
0 & 0 & 0 & 0 & 0 \\
0 & 0 & 0 & 0 & - i \\ 
0 & 0 & 0 & 0 & 0 \\
0 & 0 & 0 & 0 & 0 \\
0 & i & 0 & 0 & 0 \end{array}
\right) 
,\,\,\,
\lambda_{20} = 
\left(
\begin{array}{ccccc}
0 & 0 & 0 & 0 & 0 \\
0 & 0 & 0 & 0 & 0 \\ 
0 & 0 & 0 & 0 & 1 \\
0 & 0 & 0 & 0 & 0 \\
0 & 0 & 1 & 0 & 0 \end{array}
\right) 
,\,\,\,
\lambda_{21} = 
\left(
\begin{array}{ccccc}
0 & 0 & 0 & 0 & 0 \\
0 & 0 & 0 & 0 & 0 \\ 
0 & 0 & 0 & 0 & - i \\
0 & 0 & 0 & 0 & 0 \\
0 & 0 & i & 0 & 0 \end{array}
\right), 
\end{eqnarray}
\begin{eqnarray}
\lambda_{22}  &  = & 
\left(
\begin{array}{ccccc}
0 & 0 & 0 & 0 & 0 \\
0 & 0 & 0 & 0 & 0 \\ 
0 & 0 & 0 & 0 & 0 \\
0 & 0 & 0 & 0 & 1 \\
0 & 0 & 0 & 1 & 0 \end{array}
\right) 
,\,\,\,
\lambda_{23} = 
\left(
\begin{array}{ccccc}
0 & 0 & 0 & 0 & 0 \\
0 & 0 & 0 & 0 & 0 \\ 
0 & 0 & 0 & 0 & 0 \\
0 & 0 & 0 & 0 & - i \\
0 & 0 & 0 & i & 0 \end{array}
\right) 
,\,\,\,
\lambda_{24} = 
\frac{1}{\sqrt{10}}
\left(
\begin{array}{ccccc}
1 & 0 & 0 & 0 & 0 \\
0 & 1 & 0 & 0 & 0 \\ 
0 & 0 & 1 & 0 & 0 \\
0 & 0 & 0 & 1 & 0 \\
0 & 0 & 0 & 0 & - 4 \end{array}
\right). 
\end{eqnarray}

The definition of the scalar mesons introduced in the Hamiltonian 
 
\begin{equation}
\eta_c = \frac{1}{\sqrt{12}} ( u \bar u + d \bar d + s \bar s - 3~  c \bar c ),\,\,\,\,
\eta_b = \frac{1}{\sqrt{20}} ( u \bar u + d \bar d + s \bar s +  c \bar c - 4~ b \bar b)
\end{equation}
are consistent with the $\lambda_i$ matrices.

\vspace{1cm}

\acknowledgements

I am grateful to Ileana Guiasu for a careful reading of the manuscript.
This work has been supported by the Fonds de la Recherche Scientifique - FNRS, Belgium, 
under the grant number 4.4503.19.

\end{document}